\documentclass[twocolumn,showpacs,amsmath,amssymb,showkeys,floatfix,prl]{revtex4}
\usepackage{amsfonts,amsmath}
\usepackage{graphicx}
\usepackage{dcolumn}% Align table columns on decimal point
\usepackage{bm}% bold math

\begin{document}

\title{Quantizing the damped harmonic oscillator}

\author{D. C. Latimer}

\affiliation{Department of Physics and Astronomy, Vanderbilt University, 
Nashville, Tennessee 37235, USA}
\date{\today}

\begin{abstract}
We consider the Fermi quantization of the classical damped harmonic oscillator (dho).  
In past work on the subject, authors double the phase space of the dho in order to close the system at 
each moment in time.  For an infinite-dimensional phase space, this method requires one to construct a
representation of the CAR algebra for each time.
We show that unitary dilation of the contraction semigroup 
governing the dynamics of the system is a logical extension of the doubling procedure, and it allows
 one to avoid the mathematical difficulties encountered with the previous method.
\end{abstract}



\maketitle
\section{Introduction}
The damped harmonic oscillator (dho) is a simple classical dissipative system which, upon
quantization, yields a useful example of an open quantum system \cite{davies}.  
Unstable particles can be treated as an open system; these have been incorporated into 
quantum field theory in \cite{unstable}. 
Additionally, modeling a quantum measurement as an open system permits 
one to regard the reduction of a state as a continuous process \cite{kch}.  
Both Fermi and Bose quantization of the dho
have been considered in the literature \cite{bateman,yogi,vit1,vit3}.
A technique central to these considerations is the doubling of the degrees of freedom of the classical 
phase space, allowing one to effectively close the system for each moment in time.  The extra degrees of
freedom function as a sink with which the oscillator interacts and have been related to
quantum noise effects \cite{yogi}.  The doubled system as a whole has been expressed in the framework of
quantum deformed algebras and related to finite-temperature quantum systems
\cite{vit1,vit3}.

We shall restrict our discussion to the Fermi quantized dho, although the techniques are similar in
spirit for bosons.  After introducing the classical dho, we will discuss the doubling procedure in terms
of the unitary dilation of a contraction operator at one instant in time.  We shall relate this technique to
the representation of quasifree states over the the canonical anticommutation relation (CAR) algebra.
Following the model developed in \cite{kch}, we
introduce the unitary dilation of the contraction semigroup which describes the time evolution of the
dho.  We then discuss the quantized dho in this framework in order to demonstrate the usefulness of this
formulation.

\section{Classical dho}
A classical system can be described by a phase space $M$ of even (or infinite) dimension whose elements
specify momenta and position.  
We denote the complex structure on this space by $J$ and the projection onto the
momentum subspace as $P$.  We intend the complex structure to generate harmonic 
oscillations necessitating the additional prescription
\begin{equation}
JP = (1-P) J. \label{jpcomm}
\end{equation}
For the dho, we take the damping term to be linear in the momentum.  Damped oscillations are then
generated by the operator
\begin{equation}
Z = \omega J - 2 \gamma P,
\end{equation}
where $\omega$ is the natural (real) frequency of the oscillator and $\gamma > 0$ is the damping strength.
The dynamics of a particular point in phase space $m \in M$ are governed by the semigroup 
\begin{equation}
m(t) = T_t m, \qquad T_t = \exp [Zt], \label{semi}
\end{equation}
for $t \ge 0$.
Using the relation in (\ref{jpcomm}), we may show that the generator satisfies the quadratic
\begin{equation}
(Z + \gamma 1)^2 = (\gamma^2 - \omega^2) 1. \label{zsq}
\end{equation}
From this identity, we may write the exponential in (\ref{semi}) in a more tractable form
which explicitly demonstrates the usual behavior of a dho, depending upon the
relative values of the damping factor and the natural frequency.
Defining $\alpha = (\gamma^2 - \omega^2)^{1/2}$, we find
\begin{equation}
\exp[Zt]= e^{-\gamma t}\left[\cosh{\alpha t}1 + \frac {\sinh{\alpha t}} {\alpha}(Z + \gamma 1)
\right]. \label{exp}
\end{equation}
Should $\gamma^2 = \omega^2$, the case of critical damping, then we intend the above relation (\ref{exp}) to be
taken in the limit of vanishing $\alpha$; as such, we have
\begin{equation}
\lim_{\alpha \to 0} \frac{\sinh{\alpha t}}{\alpha} = t.
\end{equation}
An underdamped oscillator is characterized by $\gamma^2 < \omega^2$ so that $\alpha = i \omega_d$ is
a purely imaginary number.  In this case, we may write the hyperbolic functions in a more transparent form
\begin{equation}
\cosh{\alpha t} = \cos{\omega_d t}, \qquad \frac{\sinh{\alpha t}}{\alpha}= \frac{\sin{\omega_d
t}}{\omega_d};
\end{equation}
hence, one has oscillations of frequency $\omega_d$ modulated by the decaying exponential.
For an overdamped oscillator, $\gamma^2 > \omega^2$, one has real $\alpha$ so that the dynamics consist
of decaying exponentials with two different decay constants.

\section{Contraction dilation}
Given Eq.\ (\ref{exp}), 
it is apparent that $T_t$ is a contraction for all relevant
$t$ and, in fact, strongly converges to zero as $t$ tends to infinity.  
From Ref.\ \cite{sznagy}, we extract the following result concerning any contraction $T$ on $M$.
First, we define the isometric injection $j$ of $M$ into two copies of the phase space
$\widetilde{M} = M \oplus M$
\begin{equation}
jm = m \oplus 0, \qquad m \in M. \label{inject1}
\end{equation}
Then, on the doubled phase space, we may construct from the contraction the orthogonal operator
\begin{equation}
U = \left( \begin{array}{cc} T & (1- TT^*)^{1/2} \\
(1-T^* T)^{1/2} & -T^*  \end{array} \right), \label{orthog}
\end{equation}
which satisfies
\begin{equation}
j^* U j = T. \label{juj1}
\end{equation}
In words, doubling the dimension of the phase space allows one to work with an orthogonal operator,
instead of a contraction.  The additional copy of the phase space can be regarded as a sink, or coupled
oscillator.

As we are interested in the Fermi quantization of the dho, we work with the algebra 
$\mathrm{CAR}(\widetilde{M})$ where the phase space is considered 
complex with complex structure $\widetilde{J} = J \oplus -J$.  The creation operators
$c(\tilde{m})$, linear in their argument, satisfy along with their adjoints the CAR
\begin{equation}
[c(\tilde{m})^*,c(\tilde{n})]_+ = \langle
\tilde{m} ,\tilde{n} \rangle, \qquad [c(\tilde{m}),c(\tilde{n})]_+=0, \label{car}
\end{equation}
where $[\cdot,\cdot]_+$ is the anticommutator.
By definition, the operator (\ref{orthog})
generates a Bogoliubov transformation of the algebra \cite{plyrob}. 
As this is true for any contraction on $M$, in particular $T_t$, we 
conclude there is a Bogoliubov transformation associated with the dilation $U_t$ 
relating the Fock representation of
$\mathrm{CAR}(\widetilde{M})$ (at $t=0$) and the dho on the doubled phase space 
at any fixed time $t$.  This establishes connection with the previous work on the subject
\cite{bateman,yogi,vit1,vit3}.  

In general, the operator $U_t$ does not commute with the complex structure on the
doubled space.  We decompose the operator into the sum of a complex linear $a_{U_t}$ and
conjugate linear $b_{U_t}$ operators
\begin{equation}
a_{U_t} = \tfrac{1}{2}(U_t - \widetilde JU_t \widetilde J), \qquad b_{U_t} = \tfrac{1}{2}(U_t
+\widetilde J U_t 
\widetilde J).
\end{equation}
In the usual manner, we define the transformed creation operator
\begin{equation}
c_t(\tilde m) = c(a_{U_t} \tilde m) + c(b_{U_t} \tilde m)^*; \label{btc}
\end{equation}
these satisfy the CAR.
From \cite{shalestine}, we find that the Bogoliubov transformation is implementable if and only if
$b_{U_t}$ is Hilbert-Schmidt.  Given an implementable transformation, one may construct from elements within the 
Fock representation a vacuum vector that is annihilated by $c_t( \tilde m)^*$ \cite{plyrob}; in this sense, the vacuum
vector can be thought of as a dynamic object.
For a finite dimensional phase space, this situation is assured.
One may show that for fixed $t >0$, the square of the Hilbert-Schmidt norm of
$b_{U_t}$ scales with the dimension of $\widetilde{M}$. 
As such, $U_t$ is not implementable for an
infinite-dimensional phase space.  As a consequence, one must
construct a different representation for each time $t$ with no implementable means to change
between any two representations.  We remedy this situation below.

\section{Quasifree states}
First, we use similar language to elucidate the connection between this doubling procedure 
and general quasifree states over $\mathrm{CAR}(M)$.
A quasifree state $\varphi$ is determined by the two-point correlation functions.  The state can be
characterized by two bounded operators $R$ and (conjugate-linear) $S$.  These satisfy $0 \le R=R^* \le 1$ and $S^* =
-S$ with the two-point correlation functions given by 
\begin{equation}
\varphi[c(m)^* c(n)] = \langle m, R n \rangle, \qquad \varphi [c(m)c(n)] = \langle S m,  n \rangle,
\label{quasifree}
\end{equation}
for $m,n \in M$.  
The representation theory of quasifree states over the CAR algebra is well developed
\cite{verb1,araki1}; briefly, any quasifree state can be represented as a Fock state
of the CAR algebra over two copies of $M$.
For the most general quasifree state, 
the connection between its representation and the doubling procedure used with the dho is most easily accessed through
Araki's self-dual representation of the algebra \cite{araki1}.  We shall not discuss the details
of this method because the needed definitions and notation would take us too far afield.
Instead, we restrict the discussion to
quasifree states which are invariant under global
$U(1)$ phase changes; these have $S=0$, so that the state is characterized by $R$ alone. 
We are justified in restricting our scope as these are the relevant states for physics.

Given that, we note from above that $R$ is required to be a positive contraction.  As such, its
square root is defined, and $\sqrt R$ is also a contraction.  We use the injection $j$
defined in (\ref{inject1}), set the complex structure $\widetilde{J}=i \oplus -i$, and unitarily dilate 
as before
\begin{equation}
V = \left( \begin{array}{cc} \sqrt{R} & \sqrt{1- R}\\
\sqrt{1- R} & -\sqrt{R} \end{array} \right). 
\end{equation}
With this complex structure, one has
\begin{equation}
a_V = \left( \begin{array}{cc} \sqrt{R} & 0 \\
0 & -\sqrt R  \end{array} \right), \qquad 
b_V = \left( \begin{array}{cc} 0 & \sqrt{1- R} \\
\sqrt{1-R} & 0  \end{array} \right).
\end{equation}
Using transformed creation operators, as in (\ref{btc}), acting on the Fock vacuum $\Omega$, 
we calculate the two point
correlation function for elements of $M$ injected into the doubled space
\begin{eqnarray}
\langle c_V(jm) \Omega, c_V(jn) \Omega \rangle &=& \langle c(a_V jm) \Omega, c(a_V jn) \Omega \rangle \nonumber\\
&=& \langle m, R n \rangle \nonumber \\
&=& \varphi[c(m)^* c(n)],
\end{eqnarray}
as one would expect.

The parallel
with the quantized dho is particularly germane for KMS, or thermal, states.  A KMS state $\varphi_\beta$, at
inverse (positive) temperature $\beta$,
 is a quasifree state 
which describes a quantum
system with Hamiltonian $H$ in thermal equilibrium \cite{bratteli}. The defining property of this state is the
commutation relation
\begin{equation}
\varphi_\beta [A B] = \varphi_\beta [B A_{i \beta}] \label{kms}
\end{equation}
for operators $A,B$, where we have used a subscript to denote time evolution in the Heisenberg picture 
\begin{equation}
A_t = U_t^* A U_t \label{heis}
\end{equation}
with $U_t = e^{-iHt}$.
Using the KMS condition (\ref{kms}) and the CAR (\ref{car}), one can show that the state satisfies the two-point
correlation functions in (\ref{quasifree}) with $R = (1 + e^{-\beta H})^{-1}$ and $S=0$.  
Hence, we may represent the thermal
state as a Fock state on two copies of the space for each temperature $\beta > 0$, an index reminiscent of time
in the dho.  
This connection between the dho and thermal states was discussed in \cite{vit3}.  

The Fermi quantized dho and 
quasifree states share a similar mathematical structure with regard to their
representation on the doubled space.  In
particular, the analogy between the two is especially compelling for thermal states given that they are both
indexed by positive numbers (temperature and time).
However, the KMS states exhibit the much richer structure of Tomita-Takesaki theory \cite{haag}.  
As such, they necessarily have a unitary dynamical component in their definition.  
When quantizing the dho in the above manner, the dynamics of the oscillator are, in some sense, frozen out; that is,
one has a different representation at each moment in time.  Heuristically, the motivation of the doubling procedure 
is quite different for these two structures.

\section{Semigroup dilation}
Above we explicated how one may treat a single contraction as an orthogonal operator. If we are willing to inject 
the original phase space into a space even larger than $\widetilde M$, then it is possible to unitarily 
dilate the contraction semigroup $T_t$ for all $t \ge 0$.  
In what follows, we shall maintain the same notation as above in order to make clear 
that this is a logical extension of the doubling procedure.
The technique that follows can be found in \cite{sznagy}; the application of this theory to the dho was
expounded upon in \cite{kch}.
The space into which we inject $M$ is $\widehat{M}=L^2(\mathbb{R}, P M)$.  
The injection $j: M \to \widehat{M}$ is given by 
\begin{equation}
(jm)(t) = 2 \sqrt{\gamma} \Theta(t) P T_t m,
\end{equation}
with $\Theta(t)$ the Heaviside function.
This map can be shown to be an isometry.
Time translation of the elements of the space is given by the unitary operator 
\begin{equation}
(U_t \widehat{m})(s) = \widehat{m}(s+t).
\end{equation}
Analogous to Eq. (\ref{juj1}), the following holds by construction
\begin{equation}
j^* U_t j = T_t. \label{juj2}
\end{equation}

Using this technique, one no longer need bother with Bogoliubov transformations
in the above prescription or their implementation.
Though $\widehat{M}$ is a much larger space than the doubled one, the dynamics of the dho plus environment are now
unitary.
In order to keep track of the dho in this space, we employ the projection $Q=jj^*$ on $\widehat{M}$. 
One can then make the orthogonal decomposition
\begin{equation}
\widehat{M} = jM \oplus (1-Q) \widehat{M},
\end{equation}
which delineates the oscillator and the environment.
This method of dilation also exhibits a richer structure
than that of the doubling procedure. 
For instance, we may use the Fourier transform to ascertain the 
energy spread of an element of $jM$.  As energy is dual to time, we have
\begin{equation}
(\mathcal{F}jm)(E) = \frac{1}{\sqrt{2 \pi}} \int_{-\infty}^{\infty} e^{-iEt} (jm)(t) dt.
\end{equation}
After some manipulation \cite{kch}, this can be shown to be
\begin{equation}
(\mathcal{F}jm)(E) = \sqrt{\frac{2\gamma}{\pi}} \frac{1}{E^2 - \omega^2 -i2E\gamma}P(\omega J + iE) m,
\end{equation}
which has an amplitude reminiscent of the relativistic Breit-Wigner amplitude for unstable particles
(cf. \cite{peskin}).

One may quantize this total closed system in the usual manner by considering $\mathrm{CAR}(\widehat{M})$.  
The Fock representation is now adequate to describe the quantum dho.  The vacuum vector $\Omega$ is a
stationary state, and the dynamics are unitary.
We may still address the dissipative nature of the dho in this space. 
For this, we consider the second quantization $\widetilde{Q}$ of the projection $Q$
which obeys the commutation relation
\begin{equation}
[\widetilde{Q}, c(\widehat{m} )] = c(Q \widehat{m} ),
\end{equation}
for $\widehat{m} \in \widehat{M}$.  
Recalling the time dependence in the Heisenberg picture (\ref{heis}), 
we calculate the following expectation value in the Fock
representation
\begin{eqnarray}
\langle c(jm) \Omega, \widetilde{Q}(t) c(jn) \Omega \rangle &=&  \langle jm, Q(t) jn \rangle \nonumber \\
&=& \langle j^* U_t j m, j^* U_t j n \rangle \nonumber \\
&=& \langle m, T_t^* T_t n \rangle , \label{cor}
\end{eqnarray}
for $n,m \in M$.
This exhibits the behavior that one would expect. 

For the case of the critically damped oscillator, we note that
given a proper choice of initial conditions one can model a collection of unstable particles as in \cite{unstable}.  
For elements of the (nontrivial) subspace
$N=\ker(Z+\gamma 1)$, the evolution of the critically damped oscillator 
is strictly exponential, $T_t
\vert_N = e^{-\gamma t}$.  As a result, for $n \in N$, the time dependence of the 
projection onto the oscillator subspace exhibits exponential decay
\begin{equation}
\langle c(jn) \Omega, \widetilde{Q}(t) c(jn) \Omega \rangle = e^{-2\gamma t}.
\end{equation}
This is another method by which unstable particles can be included in quantum field theory.

\section{Conclusion}
In summation, we feel that the method of unitary dilation of the contraction semigroup 
is a more effective means with which to consider the quantized dho.  Rather than closing the open system at each instant in time
by doubling the dimension of the phase space, we close the entire system at once for all future time.
The advantages of this method are unitary dynamics and the need for only one Fock representation with a stationary
vacuum vector.    
The drawback for this approach is that the analogy between the dho and
thermal states is no longer valid; however, we feel that our exposition demonstrates that the two systems are
conceptually and physically disparate, lessening the significance of this shortcoming.  

\section{acknowledgements}
The author is grateful to K C Hannabuss for the mentorship he provided during the author's study 
at the Mathematical Institute, University of Oxford.

\bibliography{dho}

\end{document}